\documentclass{emulateapj}

\shorttitle{Coalescing Neutron Stars}
\shortauthors{Chaurasia \& Bailes}

\begin{document}

\title{
On the Eccentricities and Merger Rates 
of Double Neutron Star Binaries and the Creation
of ``Double Supernovae''.}


\author{H. K. Chaurasia}
\affil{Swinburne University of Technology}
\email{hemant.chaurasia@gmail.com}

\author{M. Bailes}
\affil{Swinburne University of Technology}
\email{mbailes@swin.edu.au}

\begin{abstract}

We demonstrate that a natural consequence of an asymmetric kick
imparted to neutron stars at birth is that the majority of double
neutron star binaries should possess highly eccentric orbits. This
leads to greatly accelerated orbital decay, due to the enormous
increase in the emission of gravitational radiation at periastron as
originally demonstrated by Peters (1964).  A uniform distribution of
kick velocities constrained to the orbital plane would result in
$\sim$24\% of surviving binaries coalescing at least 10,000 times
faster than an unperturbed circular system. Even if the planar kick
constraint is lifted, $\sim$6\% of bound systems still coalesce this
rapidly. In a non-negligible fraction of cases it may even be possible
that the system could coalesce within 10 years of the final
supernova, resulting in what we might term a ``double supernova''.
For systems resembling the progenitor of PSR J0737--3039A,
this number is as high as $\sim$9\% (in the planar kick model).
Whether the kick velocity distribution extends to the range required
to achieve this is still unclear. We do know that the observed
population of binary pulsars has a deficit of highly eccentric systems
at small orbital periods. In contrast, the long-period systems favour
large eccentricities, as expected. We argue that this is because the
short-period highly eccentric systems have already coalesced and are thus
selected against by pulsar surveys. This effect needs to be taken into
account when using the scale-factor method to estimate the coalescence
rate of double neutron star binaries. We therefore assert that the
coalesence rate of such binaries is underestimated by a factor of
several.

\end{abstract}


\section{Introduction}
\label{sec:intro}

Beginning with the detection of the Hulse \& Taylor double neutron star
\citep{ht75a},
pulsar astronomers have unveiled a small but growing sample of double
neutron star binaries over the last thirty years. Owing to the extreme
regularity of their radio pulse emissions, these exotic objects have
proven to be invaluable scientific laboratories, particularly for
tests of general relativity, as they are some of the few known objects
for which the effects of gravitational radiation are measurable. In fact,
the first confirmation that gravitational radiation existed was through
measurements of the orbital decay of the first known double
neutron star \citep{tw89}.

Recent surveys have started to unveil a better picture of
the double neutron star population. The Parkes multibeam 
surveys have discovered three probable double neutron star
systems \citep{lcm+00,bdp+03,fkl+05}, and Arecibo surveys have
detected two more \citep{wol91,clm+04}. Although
we are still dealing with small number statistics, the
population is now large enough to see trends emerging. 

A phenomenon of crucial importance to the binary coalescence rate is
the kick velocity imparted to a neutron star during its supernova, a
dominant factor in determining the eccentricity and coalescence time
of the subsequent orbit \citep{hil83,dc87,bai89}.  
Binary neutron star formation and evolution
also depends greatly upon this little-understood supernova kick, since
a double neutron star system must survive two such kicks in order to
remain bound. To some extent, knowledge of the kick velocity
distribution can be extracted statistically from measurements of the
observed sample of double neutron stars \citep{wk04}.
The magnitude of the kicks can be inferred by the space velocities
of pulsars, which are an order of magnitude higher than that
of their progenitors, the OB stars. Investigations of the
pulsar population demonstrate that the average kick must be
of order 400--500 km s$^{-1}$ \citep{ll94,cc98}.

Further evidence for pulsar kicks comes from precession effects observed
in the binary pulsar PSR J0045--7319 \citep{kbm+96} and 
spin-orbit coupling in the relativistic binaries PSR B1913+16,
PSR B1534+12 and PSR J1141--6545 \citep{wrt89,sta04,hbo04}.
Assuming tidal forces and mass transfer align the spin and
angular momentum axes prior to the ultimate supernova
explosion, these systems suggest a kick out of the orbital plane
has occurred. 

Binary pulsars are in some sense fossil records of the progenitor
system from which they evolved. We date these systems by the
characteristic age of the pulsar, and can therefore extrapolate
back to the orbital conditions immediately after the final supernova
explosion. What prohibits us from going further back in time 
to derive a unique progenitor is the possibility that the 
newly-born pulsar received an impulsive kick at birth. Since the
kick velocity distribution is largely unknown, including its
orientation with respect to any reference point, such as the
pulsar spin axis, we are left with a rather unsatisfactory
family of potential progenitors for any observed neutron star binary.

The relativistic nature of these objects has brought attention to the
galactic rate of double neutron star coalescence -- the merger event at
the climax of a long orbital decay and inspiral due to gravitational
radiation. Such events are some of the most significant anticipated
sources of gravitational waves, detectable by interferometers such as
the Laser Interferometer Gravity-Wave Observatory (LIGO) 
\citep{aad+92}. Double neutron star coalescence is also considered a
major candidate for explaining the origin of a subset of the mysterious
cosmic gamma-ray bursts observed by gamma-ray observatories such as
the BATSE instrument aboard the Compton Gamma-Ray Observatory
(CGRO) \citep{mfw+92}.

Much research has been conducted into predicting the double neutron
star coalescence rate, using Monte Carlo population synthesis models
\citep{bkb02,ty93,lpp97,ps96} or by statistical extrapolation
from the observed sample \citep{cl95,knst01,kkl+04}. Monte Carlo methods
are currently hampered by the vastness of the population parameter space and
compounded uncertainties, and statistical methods suffer from the
small sample size of known double neutron stars.

It is customary to estimate the event rate that advanced LIGO will
see by estimating what fraction of neutron star binaries we could
have detected in pulsar surveys if they were of similar period
and luminosity to those known, and assigning a ``scale factor'' to each.
We then multiply each member of 
the population by their scale factors and obtain an estimate of
the number of similar systems in the Galaxy, thus obtaining a
coalescence rate for the Milky Way that
can be used to infer an event rate for the local Universe by
estimating the number of Galaxies out to the LIGO detection limit.
We will refer to this method as the scale factor approach.
Whilst limited by small number statistics, 
the growing population of observed neutron
star binaries makes us increasingly confident that coalescence is
relatively common in the Universe.

The only alternative approach is to attempt to simulate the entire
binary pulsar population. Uncertainties in initial conditions
and the physics of binary evolution and kicks mean that such
estimates vary by more than two orders of magnitude \citep{bkb02}.
The population synthesis method is therefore of limited use in
estimating the merger rate.

For now, the coalescence rate is better estimated by
considering the scale factor method, however when this is done
we find the coalescence rate is dominated by the double pulsar
PSR J0737--3039A \citep{bdp+03}. This serves to demonstrate the
heavy influence of small number statistics on the uncertainties in
the scale factor method.

The scale factor method also suffers from the extent to
which the relativistic binary pulsar population remains invisible
to pulsar surveys. For instance, if there existed a population of 
neutron stars that were completely invisible in the radio we would have no
way to estimate their population from radio observations. Similarly, if only
1\% of recycled neutron stars emitted radio waves, we should
multiply the scale factors currently in use by 100. 

Our best estimates of the inspiral rate require us to think
of {\it all} the reasons why double neutron stars might be invisible 
to our surveys, and take them into account. What we explore in
this paper is a selection effect that removes binary pulsars
from our surveys and has received little attention to date.
This is the production of highly-eccentric short-period binaries with very
short lifetimes, produced via favourably-oriented kicks
in the binary progenitor. These systems are not long-lived, and
the coalescence rate of the scale-factor approach needs to be
corrected for it. Whilst this effect has been alluded to from time 
to time \citep{ty93}, we determine its importance for an arbitrary binary
in the presence of a kick velocity that can occupy any of the phase space
that will produce a bound orbit, as in this case the problem is scale invariant.

In this paper, we derive the expected distributions of orbital
eccentricity for any surviving binary pulsar, if the kick velocity
is capable of uniformly populating the available phase
space after the explosion (Section \ref{sec:ecc}).
In Sections \ref{sec:lifetimes} and \ref{sec:eccevo} we look at the lifetimes of the
surviving systems due to gravitational radiation losses 
and identify a population of binaries that coalesce very rapidly.
These results are used to explain the observed distribution of
binary pulsar eccentricities (Section \ref{sec:observedpop}), which at
long orbital periods favour high eccentricities, and at low
periods, small eccentricities. In Section \ref{sec:discussion}
we argue that the coalescence rate should be revised to
take this into consideration, particularly for the double
pulsar.

\section{Eccentricities}
\label{sec:ecc}

In this section we examine the expected orbital eccentricities of
relativistic double neutron stars from a purely analytical point of
view, avoiding any assumptions about specifics such as the
distributions of stellar masses, kick velocities and orbital
separations. Instead, we present a study of the potential
eccentricities of an arbitrary double neutron star system in units
scaled to the circular orbital speed, such that the conclusions drawn
here can be considered common to the evolution of all double neutron
star systems. Indeed, all of our results also hold for any two
compact objects (black holes, white dwarfs or neutron stars) 
where the final compact object (a black hole or a neutron star) 
is created in a
supernova and receives a kick. 

We view the effect of kick velocity through a map of the second neutron star's post-supernova
``velocity-space'' -- our representation of all possible relative velocities which correspond to bound
orbits. The boundary of this velocity-space region is circular, with radius equal to the escape speed
of the system ($\sqrt{2}$ in units of the circular orbital speed). Each point in the map corresponds
to one possible orbital velocity vector, where the position (0,1) would correspond to the circular
orbital velocity.

To calculate the semi-major axis and eccentricity of an arbitrary binary, we derive formulae from
two expressions for the total energy $E$ of the system \citep{hil83}, and two expressions for the total
angular momentum $L$ of the system \citep{md99b} :

\begin{equation}
E = -{{GMm}\over{2a}}
\label{eq:E1}
\end{equation}

\begin{equation}
E = {-{{GMm}\over{R}} + {1 \over 2}{\mu}v^2}
\label{eq:E2}
\end{equation}

\begin{equation}
L = {{\mu}vR{\rm{cos}}{\theta}}
\label{eq:L2}
\end{equation}

\begin{equation}
L = {\mu[GM_{\rm{T}}a(1-e^2)]^{1 \over 2}}
\label{eq:L1}
\end{equation}

Here $\mu=(Mm)/(M+m)$ is the binary system's reduced mass,
$M_{\rm{T}}$ its total mass, $a$ its semi-major axis, $e$ its
eccentricity, $v$ the magnitude of its relative orbital velocity
oriented at an angle $\theta$ to the plane defined by the vector that
connects the two stars, and $R$ the initial orbital separation. From
these four equations we derive expressions for the semi-major axis and
the eccentricity of an arbitrary binary, given the initial relative
orbital velocity $v$:

\begin{equation}
a = {{R}\over{2-v^2}}
\label{eq:a}
\end{equation}

\begin{equation}
e = [1-v^2{\rm{cos}}^2{\theta}(2-v^2)]^{1 \over 2}
\label{eq:e}
\end{equation}

\begin{centering}
\begin{figure}
\includegraphics[angle=270,scale=0.45]{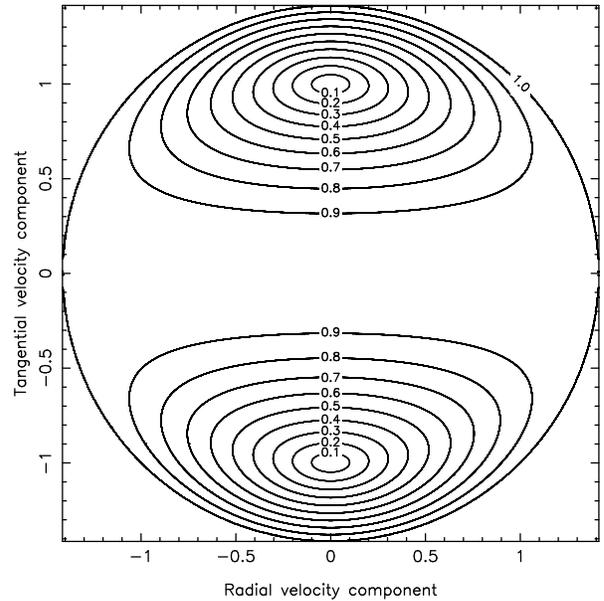}
\caption{Contour map of orbital eccentricity in the post-supernova
velocity-space of an arbitrary binary star system, in which one
star receives a randomly oriented kick. 
The velocities are measured in units of that required to place
the post-supernova system in a circular orbit -- for example, the
relative circular speed of PSR J0737--3039A is of order 600 km s$^{-1}$.
Prograde orbits have positive tangential velocities. The circular
boundary of the region represents the escape velocity of the system,
which is $\sqrt{2}$ times the circular speed.
}
\label{fig:eccmap}
\end{figure}
\end{centering}

These formulae are applied here to generate a contour map of
eccentricities in velocity-space. Note that the results are independent of
the choice of $R$.
A code was written to uniformly populate the velocity-space region
with grid points, then apply Eq. \ref{eq:e} to calculate the
eccentricity corresponding to each grid point (where each grid
point represents one possible velocity).  The 2D data was finally
plotted as a contour map, and is presented here in Fig. \ref{fig:eccmap}.

In interpreting this map, note that the relative velocity vector just
prior to supernova would be a little larger than 1.0 in the scaled
units used here, because of the larger stellar mass prior to
supernova. (For example, a pre-supernova mass of 2.0 $M_{\odot}$
corresponds to an initial orbital speed of 1.102 units.) Also, for a
sense of scale, note that the relative circular speed of PSR J0737--3039A
is of order 600 km s$^{-1}$. We operate on
the assumption that each system's final supernova is accompanied by an asymmetric
supernova kick, which has the effect of displacing the 
initial relative velocity to a new position in
velocity-space. This results in a new eccentricity and semi-major
axis, which can be calculated using the new velocity and
Eq. \ref{eq:a} and \ref{eq:e}. We further assume that the
outgoing supernova shell's impact with the companion has negligible effect.

Now, the most striking feature of the map in Fig. \ref{fig:eccmap} is
the very small size of the low eccentricity regions. This is surprising,
as the currently observed relativistic binaries tend to have
eccentricities below 0.3 and therefore must have been kicked into a
very specific region of their post-supernova velocity-space (the
third inner-most circle in Fig. \ref{fig:eccmap}). This seems an
unusually fortuitous coincidence, requiring a very consistently precise
kick. Furthermore, with mean kick magnitudes of order 400--500 km s$^{-1}$, we
expect systems to be highly prone to being kicked out of these tiny low-eccentricity
regions in velocity-space.

\begin{centering}
\begin{figure}
\includegraphics[angle=270,scale=0.35]{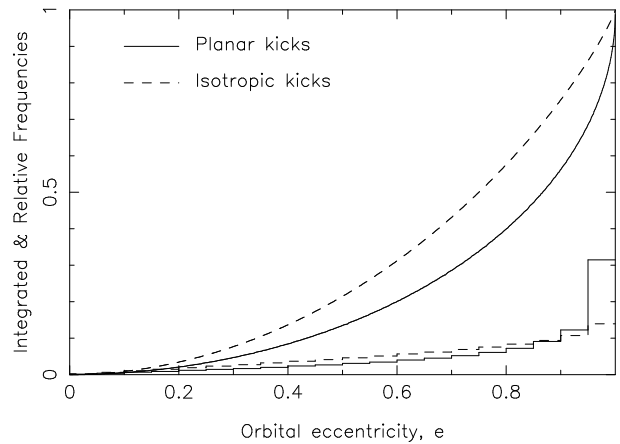}
\caption{Eccentricity distribution histograms are shown
for binaries that are produced from randomly-oriented kicks
which fill the available phase space
for the planar (solid) and general
(dashed) cases. In the general case the kicks can
be out of the plane. Also shown is the probability
that the final eccentricity will be less than
a given value, which is simply the integral of the
eccentricity distributions. Note the high probability
of eccentric systems resulting from a supernova kick
in both cases.
}
\label{fig:ecchist}
\end{figure}
\end{centering}

The 2D eccentricity data generated for Fig. \ref{fig:eccmap} is
presented as a histogram and cumulative distribution in Fig.
\ref{fig:ecchist}. Here we have also provided eccentricity data for a
3D version of our planar velocity-space, uniformly populating a
``sphere'' of radius $\sqrt{2}$. This allows us to investigate the
effect of supernova kicks which are not restricted to
the orbital plane.

The eccentricity histogram clearly illustrates a heavy bias towards
very high eccentricities in the planar velocity-space, with over 50\%
of all eccentricities greater than 0.8. The distribution is less
skewed when we allow for kick velocities out of the orbital plane, but
the bias is still towards very high eccentricities. We assert that this
illustrates a strong inherent tendency towards highly eccentric systems
arising as a consequence of supernova kicks.

This finding is not reflected in the number of highly eccentric
relativistic binaries in the observed sample of double neutron stars,
which tend to have low eccentricities.
Analysing this from the standpoint of kick velocities, the
observed systems seem to have been kicked consistently into a very
specific region of their post-supernova velocity-space, the
``bullseye'' seen in Fig. \ref{fig:eccmap}. Our explanation of this
tendency is discussed further in the following sections.

\section{Lifetimes}
\label{sec:lifetimes}

In this section we examine the probable coalescence times of
relativistic double neutron stars, using a velocity-space contour map
to illustrate dependence on the supernova kick velocity. As in the
previous section, we avoid assumptions about specifics such as the
distributions of stellar masses, kick velocities and orbital
separations, and present our results in units of the circular orbital
speed and circular coalescence time (the merger time for an arbitrary
system with zero eccentricity, denoted as $\tau_{\rm{circ}}$). These
generic units have been chosen for their usefulness in comparing
binaries in a scale-invariant manner.

\begin{centering}
\begin{figure}
\includegraphics[angle=270,scale=0.35]{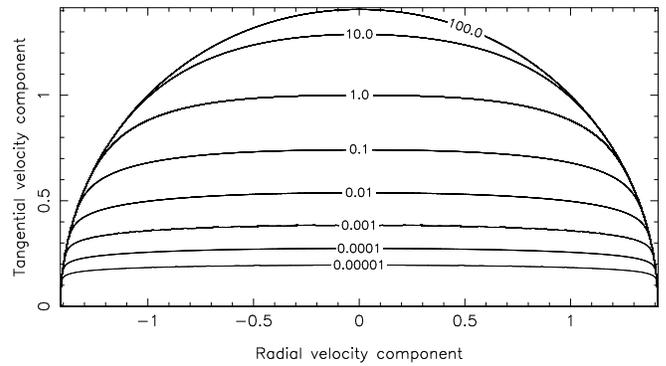}
\caption{
Contour map of coalescence times of binaries (in units of $\tau_{\rm{circ}}$)
as a function of orbital velocity immediately after
the second supernova. Only prograde orbits are shown,
but the diagram is symmetric about the $x$ axis as for
Fig 1. The time units are relative
to the coalescence time due to gravitational wave
emission for a system with a circular orbit. If neutron
stars receive a randomly oriented kick after the supernova
explosion that produces them, it is possible to populate
the entire phase space shown. Binaries with little angular momentum
after the explosion are seen here to coalesce in
remarkably short times relative to those in near-circular
orbits. The consequence of this is that such systems
are extremely difficult to observe compared to the
systems of small eccentricity.
}
\label{fig:coalmap}
\end{figure}
\end{centering}

In producing the aforementioned contour map of coalescence times, we
first generated a family of arbitrary double neutron star systems over
all possible post-supernova relative orbital velocities. Computationally,
this was achieved by uniformly populating the ``velocity-space'' with
grid-points, where each point represents a relative orbital velocity.
These velocities were then used to calculate the corresponding semi-major
axes and eccentricities, using Eq. \ref{eq:a} and \ref{eq:e}. Both of
these factors will greatly influence the rapidity of each system's orbital decay,
with close, highly eccentric binaries being much more volatile than wider,
more circular ones. To compute numerical estimates of just how rapidly
each system will coalesce, each grid point's initial orbital parameters
were evolved forward in time by numerically integrating the coupled
differential equations derived by \citet{pet64} for describing orbital
decay by relativistic gravitational radiation. A Runge-Kutta 4th order
algorithm was implemented with an adaptive step size, designed to
integrate the equations forward to the precise time when the semi-major
axis of the system drops to zero, thus determining the coalescence time.
After iterating through all grid-points, the 2D coalescence time data was
used to produce the contour map shown in Fig. \ref{fig:coalmap}.

\begin{centering}
\begin{figure}
\includegraphics[angle=270,scale=0.35]{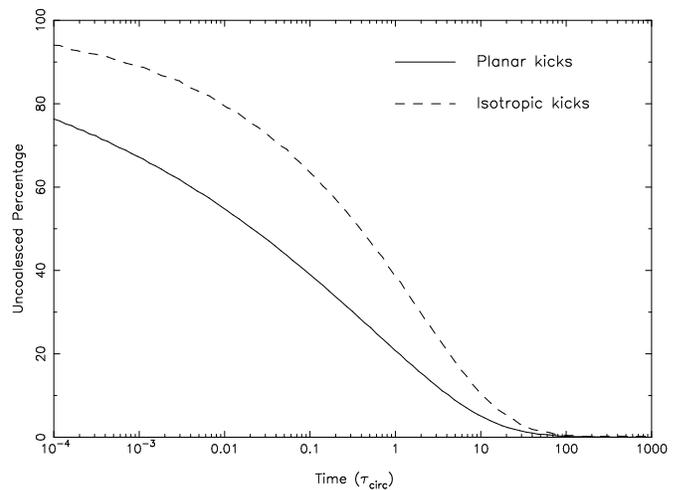}
\caption{
The fraction of systems that remain uncoalesced
as a function of time, when considering a family of
arbitrary binaries with all possible post-supernova orbital
velocities. The solid curve here illustrates the case where
the post-SN velocity is constrained to the pre-SN orbital
plane, and the dashed line represents the general case.
The unit used for time ($\tau_{\rm{circ}}$) is the coalescence
time of a system which remains in a circular orbit immediately
after the supernova. For a sense of scale, consider that
1 $\tau_{\rm{circ}}$ is equivalent to 88 Myr when scaled to the
parameters of PSR J0737--3039A/B.
}
\label{fig:pr_exist}
\end{figure}
\end{centering}

The most significant feature of Fig. \ref{fig:coalmap} is the large
central region within which coalescence times rapidly drop from 0.1
$\tau_{\rm{circ}}$ (in units of the coalescence time of a circular
system) to $10^{-5} \tau_{\rm{circ}}$ and even lower. To put this in
perspective, for the orbital parameters of PSR J0737--3039A/B, $10^{-5}
\tau_{\rm{circ}}$ corresponds to a coalescence time of less than 1,000
years. Fig. \ref{fig:coalmap} thus demonstrates that a double neutron
star system which receives a supernova kick of the order of its
circular speed has a very high likelihood of coalescing within an
incredibly short timespan.

The cumulative distribution of coalescence times is provided in
Fig. \ref{fig:pr_exist}, which shows the uncoalesced percentage of the
representative population as a function of time. This Figure dramatically
illustrates the high probability of rapidly coalescing double neutron
star systems, with up to $\sim$24\% of all systems coalescing in less
than $10^{-4} \tau_{\rm{circ}}$ (equivalent to 8,500 years when scaled
to the parameters of PSR J0737--3039). After just 0.1 $\tau_{\rm circ}$
about half of all systems have coalesced.
This is an intriguing result, implying that the
coalescence rate of double neutron stars is underestimated.
These rapidly coalescing systems
would be practically undetectable by modern pulsar surveys, since they
would not live long enough to give sufficient opportunity for
detection. They would, however, be detectable by gravitational wave
observatories such as LIGO, which will come online in the near
future. We might predict that at short orbital periods there will
be a deficit of highly eccentric systems because these are
selected against due to their shorter lifetimes. We now investigate
this in the following section.

\section{Eccentricity Distribution Evolution}
\label{sec:eccevo}

\begin{centering}
\begin{figure}
\includegraphics[angle=270,scale=0.35]{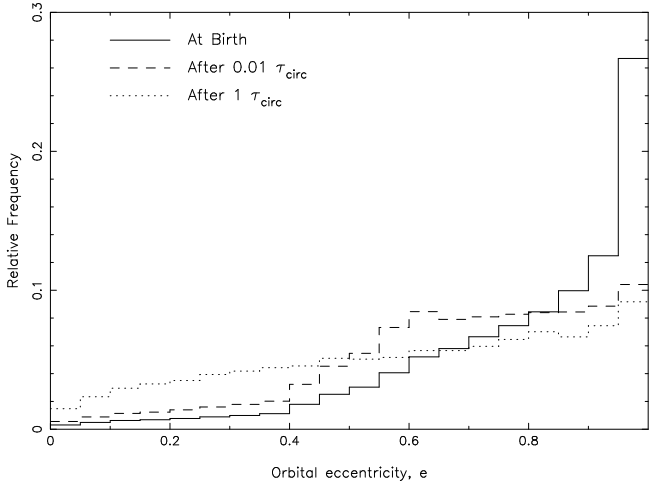}
\caption{ The evolution of observed eccentricities as a function of
time, for the binary population as in Fig. \ref{fig:ecchist}.
The units are again in terms of the time it would take a circular
system to decay due to gravitational radiation ($\tau_{\rm circ}$).
After just 0.01 $\tau_{\rm circ}$ the eccentricity distribution
(dashed) has flattened significantly as all the short-period, highly
eccentric systems have already coalesced. After one complete circular
coalesence time, the population of surviving systems (dotted) is much
more uniform.
}
\label{fig:eccevo}
\end{figure}
\end{centering}

Here we analyse the time evolution of the eccentricity distribution of
a family of arbitrary double neutron star systems. The computation
performed here is similar to that performed in the prior section -- to
begin, for every possible post-supernova relative orbital velocity
(represented computationally by uniformly spread grid-points in our
``velocity-space''), we calculate an eccentricity and semi-major axis
by applying Eq. \ref{eq:a} \& \ref{eq:e}. This results in a standard
representative family of double neutron star systems, whose orbital
parameters we can then numerically integrate forward in time using the
coupled differential equations derived by \citet{pet64}. In this way,
the eccentricity distribution of the whole family of possible double
neutron star systems can be computed at birth, and at any time
afterwards.  Fig. \ref{fig:eccevo} shows our resulting computed
eccentricity distributions at three stages in time.

This Figure serves to illustrate the rapid coalescence of the vast
majority of highly eccentric systems, which were produced in the
greatest numbers when all possible bound orbital velocities were
represented in our sample. This produces a rapidly evolving
eccentricity distribution which changes on very short timescales (0.01
$\tau_{\rm circ}$), and generally shifts towards lower
eccentricities.  While the initial post-supernova eccentricity
distribution may be highly skewed, Fig. \ref{fig:eccevo} demonstrates
that it doesn't retain its shape for very long, quickly flattening out
to a more uniform distribution, from which the observed eccentricity
distribution of double neutron star systems could be more plausibly
derived.

\section{Observed Population}
\label{sec:observedpop}

\begin{centering}
\begin{figure}
\includegraphics[angle=270,scale=0.35]{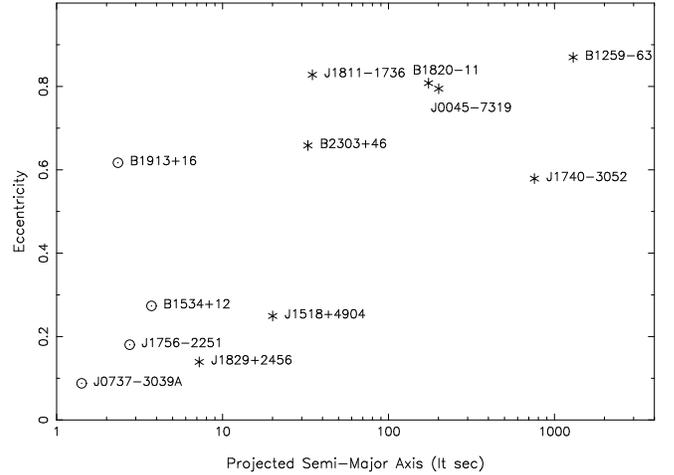}
\caption{Orbital parameters of all currently known eccentric ($e>0.05$)
binary pulsar systems residing within the disk of our galaxy.
We have excluded pulsars in globular clusters, where the formation
histories are quite different.
Relativistic systems (those which will coalesce within a Hubble time)
are indicated by dotted circles, and non-coalescing systems are
represented by stars. Note the comparatively low eccentricities of
the relativistic binaries. Pulsar information was
obtained from the ATNF Pulsar Catalogue \citep{psrcat}.
}
\label{fig:ae}
\end{figure}
\end{centering}

Here we will examine the observed population of double neutron star
binaries, particularly in terms of the eccentricity
distribution. Fig. \ref{fig:ae} shows the semi-major axes and
eccentricities of all currently known eccentric double neutron star systems
residing within the disk of our galaxy.

The majority of relativistic binaries tend to have low eccentricities,
which seems fortuitous considering the tiny size of the
low-eccentricity ``bullseye'' region in the velocity-space as shown in
Fig. \ref{fig:eccmap}.  To achieve low eccentricities, these
relativistic binary systems would have had to receive consistently
precise supernova kicks.  Any small deviation in kick direction or
magnitude would result in higher eccentricities.

Our analysis of uniformly distributed
pulsar kicks finds that surviving binaries should create mostly
highly-eccentric systems. We contend that supernova
kick velocities can easily be large enough in magnitude to kick a
binary out of its small low-eccentricity ``bullseye'' and into the
dominant, high-eccentricity region of velocity-space (see
Fig. \ref{fig:eccmap}). These large supernova kicks can be inferred
from the abundance of single pulsars roaming the galaxy at high
velocities (mean of order 400--500 km s$^{-1}$), most likely accelerated by supernova
kicks. With regards to our simple kick model, it is almost certain
that the real kick velocity distribution would not {\it uniformly}
populate the available phase space after a supernova explosion, as the kicks
in close binaries would have to range from small velocities to higher than
1000 km s$^{-1}$ with equal probability. Nevertheless, trying to guess the
exact shape and magnitude of the pulsar kick velocity distribution leads to
a plethora of models, and we resist this temptation. Our ignorance of the
kick velocity distribution does not invalidate our results, unless pulsars
can never receive a sufficiently large kick, of magnitude similar to that
of the original pre-supernova orbital velocity. We are inclined to assume
that such large kicks can exist, however, as observed pulsar velocities suggest
kicks may be imparted with magnitudes up to and exceeding 1000 km s$^{-1}$ \citep{crl93}.

We propose that the observed population of relativistic double neutron
stars exhibits low eccentricities because systems with higher
eccentricities have coalesced before they could be detected. An
average pulsar is detectable by modern pulsar surveys for
approximately 10 Myr -- however, we have shown that binary systems
have an appreciable likelihood of being kicked into coalescence within
1 Myr or even less. Not only will these systems coalesce rapidly, they
will also have orbits so tight that the effects of ``Doppler
smearing'' are too strong for current search algorithms to detect them
in some surveys. In summary, the low eccentricities of the relativistic
double neutron star systems suggest that a strong selection effect
is in operation, and we claim that accelerated decay due to gravitational
radiation is the cause.

\cite{fkl+05} suggest that there is a correlation reinforced by
the discovery of PSR J1756--2251 between orbital eccentricity and
spin period of recycled pulsars. Our results suggest that
many highly-eccentric systems resulting from close binaries are
never seen by pulsar surveys, and that long-period binaries
favour high eccentricities.
Close binaries transfer matter when the evolutionary
timescale of the donor is still long. We therefore expect that
short-period progenitors will be spun up to shorter periods
than the longer-orbital period systems.
A natural consequence of this is that the short-period
high $e$ systems are selected against, that very few long-period
low $e$ systems are born, and therefore a correlation is observed.

\section{Discussion: Implications for the Merger Rate 
and ``Double Supernovae''}
\label{sec:discussion}

We have seen that the binary pulsar population is missing
highly-eccentric binary pulsars in relativistic orbits, which we
attribute to their short observable lifetimes. The most reliable
estimates of the merger rate of neutron star binaries come from
extrapolation of the observed population, which ignores these missing
binaries that we expect to be formed in large numbers before dying
young. Our calculations suggest that factors of several are required
to correct the merger rate for these missing binaries.

We have also shown that there is a non-negligible potential for
coalescence times of the order of months if the pulsar kick is of the
same order as the pre-supernova orbital velocity. This may mean that
even relatively wide pre-supernova orbits can be driven to coalesce in
remarkably short times by receiving an appropriately-oriented
retrograde kick.  This phenomenon could reveal itself in the form of a
supernova followed closely by a second similarly-scaled catastrophic event
(the coalescence), where both would be observed at the same position a
few months or years apart -- what might be described as a ``double
supernova''.
How soon after the original burst of star formation in the Universe
might we expect to see such an event?
Population III stars --
the ancient stars which were the first to form in the history of the
Universe -- may have often occurred in binaries where both stars underwent
supernovae, subsequently taking part in an ultra-rapid coalescence which could
be detected through gravitational radiation emissions (the only known way of observing
these objects, otherwise hidden by the Epoch of Reionisation). This idea has been
alluded to before \citep{bbr04}, however our findings emphasise the possibility
that such a coalescence could occur very quickly after initial star formation, allowing
future gravitational wave detectors to directly study events that occurred within the first
few Myr of star formation in the Universe.

\acknowledgments

This paper is the product of a ten-week Summer Vacation Scholarship
undertaken by HKC at the Centre for Supercomputing and Astrophysics at
Swinburne University, Hawthorn, Australia.

\bibliographystyle{apj}

\end{document}